\documentclass[alpha-refs]{wiley-article}

\usepackage{hyperref, amsmath, color}
\usepackage[utf8]{inputenc}
\usepackage{booktabs}
\usepackage{tikz}
\usetikzlibrary{positioning}
\usepackage{indentfirst}
\usepackage{setspace}
\usepackage[normalem]{ulem}

\papertype{Original Article}

\title{Mixed effects models for healthcare longitudinal data with an informative visiting process: A Monte Carlo simulation study}

\author[1]{Alessandro Gasparini}
\author[1]{Keith R. Abrams}
\author[2]{Jessica K. Barrett}
\author[1,3]{Rupert W. Major}
\author[1,4]{Michael J. Sweeting}
\author[3,5]{Nigel J. Brunskill}
\author[1]{Michael J. Crowther}

\affil[1]{Biostatistics Research Group, Department of Health Sciences, University of Leicester, Leicester, United Kingdom}
\affil[2]{MRC Biostatistics Unit, Cambridge, United Kingdom}
\affil[3]{Department of Nephrology, University Hospitals of Leicester NHS Trust, Leicester, United Kingdom}
\affil[4]{Department of Public Health and Primary Care, University of Cambridge, Cambridge, United Kingdom}
\affil[5]{Department of Infection Immunity and Inflammation, University of Leicester, Leicester, United Kingdom}

\corraddress{Alessandro Gasparini, Biostatistics Research Group, Department of Health Sciences, University of Leicester, Centre for Medicine, University Road, Leicester, LE1 7RH, United Kingdom}
\corremail{ag475@leicester.ac.uk}

\fundinginfo{The PSP-CKD study was funded by NIHR CLAHRC East Midlands, and ongoing support is funded by NIHR CLAHRC East Midlands and Kidney Research UK. JKB is supported by the MRC Unit Programme. MJC is partially funded by the MRC-NIHR Methodology Research Panel.}

\runningauthor{Gasparini et al.}

\begin{document}

\maketitle

\begin{abstract}
  Electronic health records are being increasingly used in medical research to answer more relevant and detailed clinical questions; however, they pose new and significant methodological challenges.
  For instance, observation times are likely correlated with the underlying disease severity: patients with worse conditions utilise health care more and may have worse biomarker values recorded.
  Traditional methods for analysing longitudinal data assume independence between observation times and disease severity; yet, with healthcare data such assumptions unlikely holds.
  Through Monte Carlo simulation, we compare different analytical approaches proposed to account for an informative visiting process to assess whether they lead to unbiased results.
  Furthermore, we formalise a joint model for the observation process and the longitudinal outcome within an extended joint modelling framework.
  We illustrate our results using data from a pragmatic trial on enhanced care for individuals with chronic kidney disease, and we introduce user-friendly software that can be used to fit the joint model for the observation process and a longitudinal outcome.

\end{abstract}

\section{Introduction}\label{section:intro}

The analysis of longitudinal data is essential to understand the evolution of disease and the effect of interventions over time.
A source of longitudinally recorded data that is being used increasingly often in medical research is health care consumption data: that is, data sources that have been constructed by extracting and linking electronic health records from primary, specialist, and hospital care with other data sources such as nationwide registries for epidemiological surveillance.
Several examples of cohorts constructed in such a way are emerging in a variety of medical fields: among others, kidney disease \citep{hemmelgarn_2009, runesson_2015}, cardiovascular disease \citep{denaxas_2012}, and end-of-life healthcare \citep{tanuseputro_2015}.
Data cohorts constructed by extracting medical records have thousands - if not millions - of individuals with hundreds of measurements each: the availability to researchers of such vast amount of data allows answering more relevant and detailed clinical questions but poses new challenges.
In terms of reporting, guidelines have emerged to improve discovery, transparency, and replicability of research finding utilising routinely collected data \citep{benchimol_2015}.
In terms of methodological challenges, first and foremost, observation times are likely to be correlated with the underlying disease severity in healthcare consumption datasets.
For instance, individuals tend to have irregular observation times as patients with more severe conditions (or showing early symptoms of a disease) tend to visit their doctor or go to the hospital more often than those with milder conditions (and no symptoms).
Their worse disease status is also likely to be reflected in worse biomarkers being recorded at such visits, causing abnormal values of such biomarkers to be overrepresented and normal values to be under-represented.
Taking this pattern to the extreme, healthy individuals may not appear in health records at all leading to cohort selection bias; this is a separate issue that is not dealt with in this manuscript.

Traditional methods used to analyse longitudinal data rely on the assumption that the underlying mechanism that controls the observation time is independent of disease severity; however, that is unlikely with health care consumption data.
It can be shown that failing to account for informative dropout in a longitudinal study could yield biased estimates of the model parameters \citep{wu_1988, mcculloch_2016}, and so does n\"aively applying traditional methods when the follow-up is irregular and related to the outcome \citep{pullenayegum_2016}.
Despite the potential for bias, there is some evidence pointing towards a lack of awareness of the potential for bias in longitudinal studies with healthcare data irregularly collected over time: in a recent literature review on the topic, \citet{farzanfar_2017} showed that \(86\%\) of studies did not report enough information to evaluate whether the visiting process was informative or not, and only one study used a method capable of dealing with an informative observation process.
This is concerning when the aim of a research project is aetiology.

Bias may arise when data on covariates and outcomes is collected at irregular, subject-specific intervals: in fact, when analysing data originating from electronic health records, data is collected only when study subjects consume health care (e.g. by visiting their doctor or going to the hospital).
As a consequence, visit times are likely to be informative and to depend on the clinical history of an individual.
The visiting process in this setting is therefore deemed to be informative (or dynamic, outcome-dependent).
The bias that one may encounter when the observation process is informative can be classified in two types: selection bias or confounding \citep{hernan_2009}.
Selection bias arises because of the selection of observed individuals only in the analysis.
This bias is the same bias induced by informative censoring due to loss to follow-up \citep{hernan_2004}: censoring is the extreme case of an observation process where an individual is not observed ever again.
Conversely, confounding arises when there are common causes of both the exposure and the outcome, e.g. when the consequent visit times are decided by physicians or patients based on e.g. current health status, which itself is associated with the observed longitudinal outcome.
\citet{hernan_2009} describe selection bias and confounding originating from dynamic observation processes more in detail, including directed acyclic graphs (DAGs) that illustrate the underlying causal mechanism.

In the past years several methods have been developed to deal with longitudinal data terminated by informative dropout \citep{kurland_2009}; conversely, the problem of informative visit times has received considerably less attention.
Despite that, a few methods emerged that can be broadly categorised in two families: methods based on inverse intensity of visit weighting (IIVW, an extension of inverse probability of treatment weighting, \citet{robins_1995}) and methods based on shared random effects \citep{liu_2008a}.
An introduction to the various methods is presented elsewhere \citep{pullenayegum_2016}.
Nevertheless, to the best of our knowledge, there is only one comparison existing in the current literature, which yielded negative results: \citet{neuhaus_2018} conclude that fitting ordinary linear mixed models disregarding the observation process yielded the smallest bias and showed that adding regular visits to the observation schedule (if possible) reduced that bias even further.

Throughout this paper, we focus on the problem of informative visiting process by assuming that the dropout process is not informative.
First, we describe characteristics of the observation process and we define when it can be deemed informative in Section \ref{section:obsprocess}.
Then, we introduce a joint model for the observation and longitudinal processes that can be easily extended within a multivariate generalised linear and non-linear mixed effects models framework \citep{crowther_2017} in Section \ref{section:model}, and introduce the IIVW method in more detail in Section \ref{section:iivw}.
We compare the performance of this model against other alternatives that have been introduced in the literature via Monte Carlo simulation in Section \ref{section:simulation}.
Finally, we illustrate the use of the joint model using data from a pragmatic trial in chronic kidney disease and discuss our conclusions in Sections \ref{section:application} and \ref{section:discussion}, respectively.

\section{Characteristics of the observation process}\label{section:obsprocess}

An observation process can have regular or irregular visits.
With regular visits, the j\textsuperscript{th} visit time for the i\textsuperscript{th} individual \(T_{ij}\) is the same for all individuals: \(T_{ij} = t_j \ \forall \ i, \ j\),  with \(i = 1, 2, \dots, n\) and \(j = 1, 2, \dots, n_i\).
Conversely, with irregular visits that is no longer true.
With irregular visits, the observation process - denoted by the counting process \(N_i(t)\) - can be defined to be completely at random when visit times and outcomes are independent \citep{pullenayegum_2016}:
\[
E[\Delta N_i(t) | \bar{Y}_i(\infty), \bar{X}_i(\infty)] = E[\Delta N_i(t)],
\]
where \(\Delta N_i(t) = N_i(t) - N_i(t^{-})\), and \(t^{-}\) being the instant of time right before \(t\).
\(\bar{Y}_i(\infty)\) and \(\bar{X}_i(\infty)\) denote the values of outcome and covariates for any \(t > 0\).

The observation process can be deemed informative when it is not completely at random, i.e. when the condition above is not verified.
In that case, it is possible to identify the following two scenarios:
\begin{itemize}
    \item Observation process at random, when visiting at time \(t\) is independent of the outcome at time \(t\) given data recorded up to time \(t\):
                \[
                    E[\Delta N_i(t) | \bar{X}_i(t), \bar{N}_i(t^{-}), \bar{Y}_i^{\text{obs}}(t^{-}), Y_i(t)] = E[\Delta N_i(t) | \bar{X}_i^{\text{obs}}(t), \bar{N}_i(t^{-}), \bar{Y}_i^{\text{obs}}(t^{-})],
                \]
                where \(\bar{X}_i(t)\) and \(\bar{X}_i^{\text{obs}}(t)\) denote the covariates history up to time \(t\) and its observed values, \(\bar{N}_i(t^{-})\) denotes the history of the observation process up to time \(t^{-}\), and \(\bar{Y}_i^{\text{obs}}(t^{-})\) the observed values of the outcome up to time \(t^{-}\);
    \item Observation process not at random, where the definition of missing at random does not hold. That is, the scenario where visiting at time \(t\) is not independent of the outcome at time \(t\), even after conditioning on data recorded up to time \(t\):
                \[
                    E[\Delta N_i(t) | \bar{X}_i(t), \bar{N}_i(t^{-}), \bar{Y}_i^{\text{obs}}(t^{-}), Y_i(t)] \neq E[\Delta N_i(t) | \bar{X}_i^{\text{obs}}(t), \bar{N}_i(t^{-}), \bar{Y}_i^{\text{obs}}(t^{-})]
                \]
\end{itemize}

\cite{gruger_1991} illustrate four possible models that could be linked to the above-mentioned scenarios:
\begin{enumerate}
    \item the \emph{examination at regular intervals} model, consisting of observation times that are pre-defined and equal for all patients. This scenario yields so-called \emph{balanced panel data};
    \item the \emph{random sampling} model, consisting of a sampling scheme (e.g. an observation process) that is not pre-defined, but still independent of the disease history of the study subjects;
    \item the \emph{doctor's care} model, consisting of an observation process that depends on the characteristics of the patient at the moment of the current doctor's examination. For instance, a doctor could require stricter monitoring for subjects with more advanced disease status, or with abnormal values of a biomarker;
    \item the \emph{patient self-selection} model, yielding observations that are triggered by the patients themselves. According to this model, patients may choose to visit their doctor when they feel unwell, or they may choose to skip a visit that was pre-planned when they feel the treatment they are receiving is not beneficial to their health status. Unfortunately, the factors that cause patients to self-select themselves are generally unknown or not recorded.
\end{enumerate}

Models (1) and (2) could be characterised as \emph{observation completely at random}; model (3) could be characterised as \emph{observation at random}; finally, model (4) could be characterised as \emph{observation not at random}.

\section{A joint model for the observation process and a longitudinal outcome}\label{section:model}

Let \(D_{ij}(t) = I(T_{ij} = t)\) denote the presence of an observation at time \(t\) for the \(i\)\textsuperscript{th} individual: at each \(D_{ij}(t) = 1\) a new observation of the longitudinal outcome \(Y_{ij}\) is recorded.
Let \(\tilde{t}_{ij}\) be the gap time between the \(j\)\textsuperscript{th} and \((j+1)\)\textsuperscript{th} measurement for the \(i\)\textsuperscript{th} individual.
Let \(\tilde{d}_{ij}\) be the binary indicator variable that denotes whether the gap-time \(\tilde{t}_{ij}\) is observed (or not).
In practice, gap-time are always observed except when the observation process is censored at the end of follow-up, e.g. the date when the data extraction occurs.
Let \(z_{ij}\) be the covariate vector for the longitudinal outcome, and \(w_i\) the covariate vector for the observation process; \(z_{ij}\) and \(w_i\) do not necessarily overlap, and it is assumed that both could be extended to include time-dependent exogenous covariates (e.g. \(w_{ij}\)).
We model the observation process and the repeated measures process using a joint longitudinal and survival model.
Conditional on random effects \(u_i\), the submodel for the time to each observation is a proportional hazards model with hazard for gap time \(\tilde{t}_{ij}\):
\[
r(\tilde{t}_{ij} | w_{ij}, u_i, \theta_t) = r_0(\tilde{t}_{ij}) \exp(w_{ij} \beta + u_i), \tag{1}
\]
where \(\theta_t = \beta\).
The submodel for the j\textsuperscript{th} longitudinal observation for the i\textsuperscript{th} individual is
\[
(y_{ij} | D_{ij}(t) = 1, z_{ij}, u_i, v_i, \theta_y) = m_{ij} + \epsilon_{ij} = z_{ij} \alpha + \gamma u_i + v_i + \epsilon_{ij}, \tag{2}
\]
where \(\epsilon_{ij} \sim N(0, \sigma^2_{\epsilon}) \) and \(\theta_y = \{\alpha, \gamma, \sigma^2_{\epsilon}\}\).

Equation 1 is a recurrent events model for the observation process, with \(r_0(\tilde{t}_{ij})\) any parametric or flexible parametric \citep{royston_2002} baseline hazard function (also referred to as baseline intensity - we use the terms hazard and intensity interchangeably throughout this manuscript).
Equation 2 is a linear mixed model for the longitudinal outcome with a random intercept \(v_i\).
The two processes are linked together via the shared, individual-specific, random effect \(u_i\).
Including the \(\gamma\) parameter in the longitudinal model allows for an association between the two equations, association that will be estimated from data; when \(\gamma = 0\), the two processes are independent of each other; that is, the observation process is not informative.
Finally, we assume that the random effects follow a multivariate normal distribution with null mean vector and variance-covariance matrix \(\Sigma_{u,v}\).

The model is fitted using maximum likelihood; the individual-specific contribution to the likelihood can be written as:
\begin{align*}
L_i(\theta) &= \int p(\tilde{t}_{ij}, \tilde{d}_{ij}, y_{ij}, b_i; \theta) \ d b_i \\
            &= \int \prod_{j = 1}^{n_i} p(\tilde{t}_{ij}, \tilde{d}_{ij} | b_i, \theta_t) p(y_{ij} | b_i, \theta_y) p(b_i | \theta_b) \ d b_i
\end{align*}
where \(\theta = \{\theta_t, \theta_y, \theta_b\}\) is the overall parameters vector, \(b_i = \{u_i, v_i\}\) is the vector of random effects,
\[
p(\tilde{t}_{ij}, \tilde{d}_{ij} | b_i, \theta_t) = r(\tilde{t}_{ij} | w_{ij}, u_i, \theta_t) ^ {\tilde{d}_{ij}} \exp \left( -\int_0^{\tilde{t}_{ij}} r(s | w_{ij}, u_i, \theta_t) \ ds \right)
\]
is the contribution to the likelihood of the time to the j\textsuperscript{th} observation in individual i,
\[
p(y_{ij} | b_i, \theta_y) = (2 \pi \sigma^2_{\epsilon}) ^ {-1/2} \exp \left( - \frac{(y_{ij} - m_{ij})^2}{2 \sigma^2_{\epsilon}} \right)
\]
is the contribution of the j\textsuperscript{th} longitudinal observation, and \(p(b_i | \theta_b)\) is the density of the random effects.
The likelihood does not have a closed form, as it is necessary to integrate out the distribution of the random effects; methods such as Gaussian quadrature and Monte Carlo integration can be used for that purpose \citep{pinheiro_1995}.

\begin{figure}
\centering
	\begin{tikzpicture}
		\node (1) {Y};
		\node[right = of 1] (2) {X};
		\node[below = of 2] (3) {R};
		\node[below = of 1, rectangle, draw] (4) {U};
		\draw [->] (2) -- (1);
		\draw [->] (2) -- (3);
		\draw [->] (4) -- (1);
		\draw [->] (4) -- (3);
	\end{tikzpicture}
	\caption{Simplified DAG depicting a joint model for a longitudinal outcome and its observation process.}
	\label{fig:jm-dag}
\end{figure}
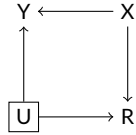

A simplified DAG that illustrate how the joint model accounts for the correlation between a longitudinal outcome \(Y\) and its observation process \(R\) is included as Figure \ref{fig:jm-dag} \citep{liu_2018}; \(X\) represents covariates included in the model, and \(U\) represents the shared random effects.
After adjusting for all covariates (e.g. confounders) \(X\), the longitudinal outcome and the observation process are associated only through the shared \(U\).
However, when estimating the joint model we assume a distribution for \(U\) (e.g. Gaussian) and we integrate it out of the marginal likelihood, blocking the path between \(Y\) and \(R\).
Therefore, for the joint model to be valid the observation process has to be at least \emph{at random}, according to the definition of Section \ref{section:obsprocess}.

This model is nested within a wide family of multivariate generalised linear and non-linear mixed effects models \citep{crowther_2017}. The model presented in this section can easily be extended to multiple random effects (potentially nested within each other), to different parametric and flexible parametric baseline hazard formulations for the recurrent events model, and to include other outcomes (e.g. a dropout process, or a second longitudinal outcome); we focus on the model formulated in this section for simplicity. Finally, this joint model (and several extensions) can be easily fitted in Stata using the user-written command \texttt{merlin} \citep{crowther_2018}. We produce example code that is included in the Online Supplementary Material.

\section{Inverse intensity of visit weighting}\label{section:iivw}

The bias induced by an informative observation process can be adjusted for by using the inverse intensity of visit weighting [IIVW] method first proposed by \citet{robins_1995} as an extension of the inverse probability of treatment [IPW] method \citep{cole_2008}.
This method was further developed by \citet{buzkova_2007}, and there are a few examples of this method applied in practice \citep{van_ness_2009, buzkova_2010}.
The IIVW approach accommodates an informative observation process in a marginal regression model by weighting each observation by the inverse of the probability of each measurement to be recorded.
This approach creates a pseudo-population in which the observation process is static and can be ignored.
The weights can be estimated by fitting a regression model including all covariates that inform the observation process and further stabilised to increase efficiency \citep{cole_2008}.
The weighting model could include current and past values of any covariate that may affect the visiting process; however, as with IPW, all covariates that might be related to the observation process should be included in the weighting model - otherwise, bias will incur.

The approach we illustrate follows from \citet{van_ness_2009}.
The model used to estimate weights is an Andersen-Gill recurrent events model \citep{andersen_1982} for the observation process, assuming a gap-time scale (as described in Section \ref{section:model}):
\[
r(\tilde{t}_{ij}) = r_0(\tilde{t}_{ij}) \exp(z_i \eta),
\]
where \(\tilde{t}\) are gap-times between consecutive observations, \(r_i(\tilde{t})\) is the intensity of visit for individual \(i\) at gap-time \(\tilde{t}\), \(r_0(\tilde{t})\) is the unspecified baseline intensity at gap-time \(\tilde{t}\), and \(z_i\) is a vector of coefficients that are assumed to accurately describe the observation process for individual \(i\).
\(\eta\) is a vector of regression coefficients that is estimated using the Cox partial likelihood method and a robust jack-knife estimator for the variance of the regression coefficients.
The inverse intensity of visit weights are estimated by taking the inverse of the linear predictor \(\exp(z_i \hat{\eta})\) at each time point, and further normalised by subtracting the mean inverse weight and adding the value 1 to each weight; the distribution of the weights is therefore centred on the value 1.
Finally, two further adjustments are needed.
First, since the last data entry for each individual represents the end of follow-up of the study, each weight is shifted by one time point.
Second, given that each individual is observed at least once (i.e. at baseline), a weight of one is assigned to the first observation of each individual.

The marginal model for the longitudinal outcome is then fit using generalised estimating equations and including the normalised inverse intensity of visit weights as probability weights in the model.
The model has the form
\[
E(y_{ij}) = \alpha_0 + Z_i \alpha_1 + t_{ij} \alpha_2,
\]
and can be fit using readily available statistical software.
We use the Stata command \texttt{glm}.

\section{A Monte Carlo simulation study}\label{section:simulation}

\paragraph{Aim}
We design a simulation study aimed to assess the impact of ignoring the observation process in longitudinal mixed-effects models when the observation process is informative.

\paragraph{Data-generating mechanisms}
We simulate data from the following joint model:
\begin{align*}
r(\tilde{t}) &= r_0(\tilde{t}) \exp(Z_i \beta + u_i) \\
y_{ij} | (D_{ij}(t) = 1) &= \alpha_0 + Z_i \alpha_1 + t_{ij} \alpha_2 + \gamma u_i + v_i + \epsilon_{ij}
\end{align*}
\(Z_i\) is a time-invariant covariate (for simplicity) representing a binary treatment, simulated from a Bernoulli random variable with probability \(0.5\): \(Z_i \sim \operatorname{Bern}(1, 0.5)\).
The coefficient associated to the treatment variable is \(\beta = 1\) for the observation process, \(\alpha_1 = 1\) for the longitudinal process.
The fixed intercept of the longitudinal model is \(\alpha_0 = 0\), and the fixed effect of time is \(\alpha_2 = 0.2\).
The random effects \(u_i\) and \(v_i\) are simulated from a Normal random variable with null mean and variance \(\sigma_u^2 = 1\) and \(\sigma_v^2 = 0.5\), respectively.
The residual error of the longitudinal model is assumed to follow a Normal distribution with null mean and variance \(\sigma_{\epsilon}^2 = 1\).
We assume independence between the random effects and the residual variance, and between random effects (i.e. \(\Sigma_{u,v}\) is a diagonal matrix with \(\text{diag}(\Sigma_{u,v}) = \{\sigma_u^2, \sigma_v^2\}\)).
We assume independent random effects for simplicity, but we show in the Online Supplementary Material how to fit a joint model with correlated random effects.
The joint model with correlated random effects can be thought of as a reparameterisation of the joint model with independent random effects, where the association parameter \(\gamma\) is related to the correlation between the two random effects in the bivariate version.
The baseline hazard from the recurrent visit process is assumed to follow a Weibull distribution with shape parameter \(p = 1.05\); we vary the scale parameter \(\lambda\) and therefore the baseline intensity of the visiting process, with \(\lambda = \{0.10, 0.30, 1.00\}\).
This baseline intensities along with the value of \(\beta\) correspond to an expected median gap time between observations of 5.83 and 2.25 years for unexposed and exposed individuals if \(\lambda = 0.10\), 2.05 and 0.79 years if \(\lambda = 0.30\), and 0.65 and 0.25 years if \(\lambda = 1.00\), respectively.
Each observation time is simulated using the inversion method of \citet{bender_2005}, assuming a gap time scale (where the time index is reset to zero after the occurrence of each observation; the resulting recurrent events model is then a semi-Markov model).
We vary the association parameter \(\gamma\) between the two sub models, with \(\gamma = \{0.00, 1.50\}\); we expect all models to perform similarly when \(\gamma = 0\), that is, when the longitudinal process is independent of the observation process.

In addition to simulating data from the joint model above, we generate the observation process by drawing from a Gamma distribution.
Specifically, we draw the observation times from a Gamma distribution with shape = 2.00 and scale:
\[
\exp(-\psi \beta Z_i + \xi_i),
\]
where \(\xi_i\) is simulated from a Normal distribution with null mean and variance \(\sigma_{\xi}^2 = 0.1\).
\(Z_i\) is the same binary treatment covariate as before, with the same associated parameter \(\beta = 1\).
The value of \(\psi\) defines the association between the observation, e.g. when \(\psi = 0\) the observation process is not informative; we set \(\psi = \{0.00, 2.00\}\).
We also simulate a scenario where the observation process depends on treatment and on previous values of the longitudinal outcome \(Y\).
In this setting, we draw observation times from a Gamma distribution with shape = 2.00 and scale:
\[
\exp(-\psi \beta Z_i + \omega y_{i,j-1} + \xi_i)
\]
for the j\textsuperscript{th} observation time of the i\textsuperscript{th} individual, with \(\psi = 2.00\) and \(\omega = 0.20\).
Finally, we simulate a scenario from a joint model to which we add regular (i.e. planned) visits every year, as suggested by \citet{neuhaus_2018}.
We simulate this scenario from the above-mentioned joint model, and we set \(\gamma = 3.00\) and \(\lambda = 0.05\) to obtain an observation process that is sparse and strongly associated with the longitudinal outcome.

We simulate 200 study individuals under each data-generating mechanism and the recurrent observation process continues for each individual until the occurrence of administrative censoring - which we simulated from a \(\operatorname{Unif}(5, 10)\) random variable.

We define the last gap time for each individual as the difference between the last observation and the censoring time.

\paragraph{Estimands}
The main estimand of interest is the vector of regression coefficients \(\alpha = \{\alpha_0, \alpha_1, \alpha_2\}\), with specific focus on the treatment effect \(\alpha_1\).
In the Online Supplementary Material, we also report on the estimated association parameter \(\gamma\) and on the estimated variance of the random effects and the residual errors: \(\sigma_u^2\), \(\sigma_v^2\), and \(\sigma_{\epsilon}^2\).

\paragraph{Methods} We fit five competing models to each simulated dataset:
\begin{enumerate}
  \item model A, the joint model described above (at the beginning of the "Data-generating mechanisms" section) and corresponding to the true data-generating mechanisms when simulating data from a joint model;
  \item model B, a linear mixed model including the number of visits (centred on the mean value) as a fixed effect in the model;
  \item model C, a linear mixed model including the cumulative number of visits as a fixed effect in the model;
  \item model D, a linear mixed model that disregards the observation process completely;
  \item model E, a marginal model fitted using generalised estimating equations and inverse intensity of visit weights.
\end{enumerate}

Model A is fit using \texttt{merlin} \citep{crowther_2018} and \texttt{gsem} in Stata.
Model B follows from previous work by \citet{goldstein_2016}, where they demonstrate that conditioning on the number of health-care encounters it is possible to remove bias due to an informative observation process (they denote this bias as "informed presence bias").
We therefore include the number of observations per individual, centred on the mean value, in a mixed effects model for the longitudinal outcome:
\[
y_{ij} = \alpha_0 + Z_i \alpha_1 + t_{ij} \alpha_2 + n^c_i \alpha_3 + v_i + \epsilon_{ij},
\]
with \(v_i\) a random intercept and \(n^c_i\) the number of observations for the \(i\)\textsuperscript{th} individual.
Model C is analogous to model B, adjusting for the cumulative number of measurements up to time \(j\) instead, denoted as \(\bar{n}_{it_{j}}\):
\[
y_{ij} = \alpha_0 + Z_i \alpha_1 + t_{ij} \alpha_2 + \bar{n}_{it_{j}} \alpha_3 + v_i + \epsilon_{ij}
\]
Model D is analogous to model B and C, assuming \(\alpha_3 = 0\).
Model B, C, and D are fit using the \texttt{mixed} command in Stata.
Model A, B, C, D are fit assuming an independent structure for the variance-covariance matrix of the random effects.
Finally, model E is fitted following the two-stage procedure presented in \citet{van_ness_2009} and illustrated in Section \ref{section:iivw}.

\paragraph{Performance measures}
We will assess average estimates and standard errors, empirical standard errors, bias, and coverage probability of \(\hat{\alpha}_m\), with \(m = \{0, 1, 2\}\).
However, the main performance measures of interest are bias and coverage probability: the former quantifies whether an estimator targets the true value on average, while the latter represents the proportion of times that a confidence interval based on \(\hat{\alpha}_{m,k}\) and \(\hat{\text{SE}}(\hat{\alpha}_{m,k})\) contains the true value \(\alpha_m\), with \(k\) indexing each replication.
We compute and report Monte Carlo standard errors to quantify the uncertainty in estimating bias and coverage \citep{morris_2019}.
If we assume that \(\text{Var}(\hat{\alpha}_m)\)\(\le 0.1\) (or, equivalently, \(\text{SE}(\hat{\alpha}_m)\)\(\le 0.32\)) and we require a Monte Carlo standard error for bias of \(0.01\) or lower, given that \(\text{MCSE}(\text{Bias}) =\) \(\sqrt{\text{Var}(\hat{\alpha}_m) / K}\), we would require a number of replications K = 1,000.
The assumed standard error is larger than the standard errors reported by \citet{liu_2008a} for a model similar to model A.
The expected Monte Carlo standard error for coverage, assuming a worst-case scenario of \(\text{coverage} = 0.50\), would be \(0.02\) - which we deem acceptable.
Therefore, we proceed by simulating 1,000 independent data sets for this simulation study.

\paragraph{Software}
The simulation study is coded and run using Stata version 15, built-in functions (such as \texttt{mixed}, \texttt{glm}, \texttt{gsem}), and the user-written commands \texttt{survsim} \citep{crowther_2012a} and \texttt{merlin} \citep{crowther_2018}; results of the simulation study are summarised using R \citep{Rmanual} and the R package \texttt{rsimsum} \citep{gasparini_2018}.
All the code required to simulate data, fit each model, and produce summary tables and figures is publicly available on the GitHub page of the first author (\url{https://github.com/ellessenne/infobsmcsim}).

\begin{figure}
  \centering
  \includegraphics[width = 0.90\textwidth]{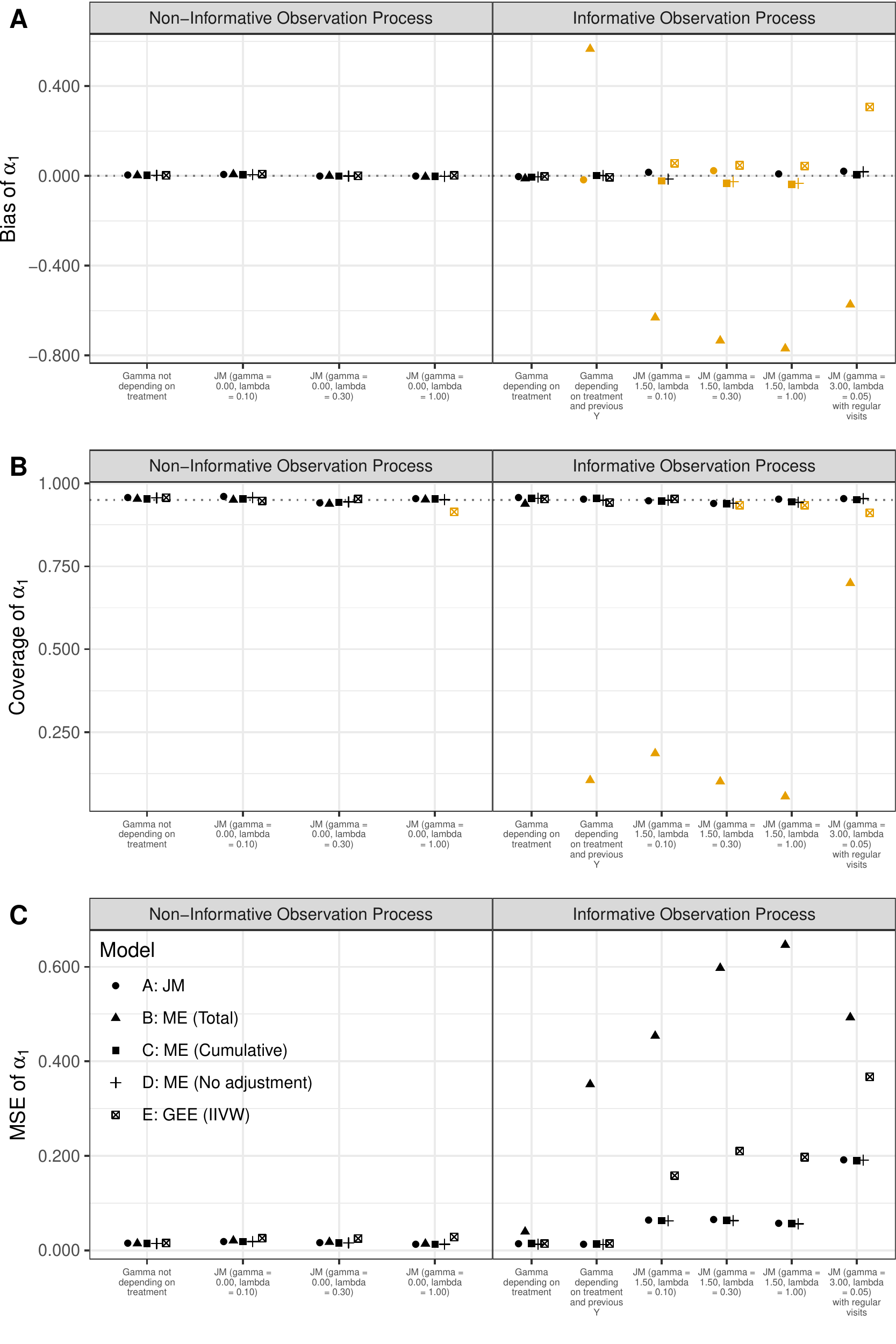}
  \caption{Bias (panel A), coverage (panel B), and mean squared error (panel C) of the estimated treatment effect \(\alpha_1\). The orange colour identifies scenarios where the summary statistics was significantly different than the target value (0 for bias, 95\% for coverage) using Z-tests based on estimated Monte Carlo standard errors.}
  \label{fig:alpha1}
\end{figure}

\paragraph{Results}
We focus on results for the estimated treatment effect \(\alpha_1\), which are depicted in Figure \ref{fig:alpha1}.
Tabulated values are included in the Online Supplementary Material, alongside results for the other regression coefficients \(\alpha_0, \alpha_2\), estimated variances of the random effects, summaries for the association parameter \(\gamma\), and convergence rates of each model under each data-generating mechanism.

\subparagraph{Descriptive results}
Each simulated dataset had 200 distinct individuals; summary descriptive statistics for each data-generating mechanism are included in Table \ref{tab:dgmsumm}.
The median sample size per simulated dataset varied between 666 and 4,482, the median number of measurements per individual varied between 2 and 13, and the median gap time between observations varied between 0.13 and 1.31 (years).
In simulated scenarios from the joint model, as expected, a higher baseline intensity of visit process yielded more frequent measurements and a larger number of measurements overall; values were not affected by the association parameter.

\begin{table}[!h]

\caption{\label{tab:dgmsumm}Summary characteristics of simulated data under each data-generating mechanism. Values are median with inter-quartile interval [IQI].}
\centering
\resizebox{\linewidth}{!}{
\begin{tabular}{rrrr}
\toprule
Data-Generating Mechanism & Sample Size & N. of Measurements & Gap Time\\
\midrule
\(\Gamma\) distribution not depending on treatment & 938 (918 - 957) & 4 (3 - 6) & 1.31 (0.74 - 2.17)\\
JM (\(\gamma = 0.00, \lambda = 0.10\)) & 666 (634 - 705) & 2 (1 - 4) & 0.91 (0.33 - 2.12)\\
JM (\(\gamma = 0.00, \lambda = 0.30\)) & 1,564 (1,475 - 1,667) & 5 (2 - 9) & 0.37 (0.13 - 0.94)\\
JM (\(\gamma = 0.00, \lambda = 1.00\)) & 4,489 (4,188 - 4,815) & 13 (6 - 27) & 0.13 (0.04 - 0.33)\\
\(\Gamma\) distribution depending on treatment & 3,444 (3,296 - 3,606) & 11 (4 - 28) & 0.23 (0.12 - 0.41)\\
\(\Gamma\) distribution depending on treatment and previous Y & 2,564 (2,457 - 2,670) & 9 (4 - 20) & 0.31 (0.16 - 0.60)\\
JM (\(\gamma = 1.50, \lambda = 0.10\)) & 669 (637 - 707) & 2 (1 - 4) & 0.90 (0.33 - 2.11)\\
JM (\(\gamma = 1.50, \lambda = 0.30\)) & 1,556 (1,461 - 1,654) & 5 (2 - 9) & 0.37 (0.13 - 0.94)\\
JM (\(\gamma = 1.50, \lambda = 1.00\)) & 4,482 (4,218 - 4,794) & 13 (6 - 26) & 0.13 (0.04 - 0.33)\\
JM (\(\gamma = 3.00, \lambda = 0.05\)) with regular visits & 1,842 (1,818 - 1,867) & 9 (7 - 10) & 1.00 (1.00 - 1.00)\\
\bottomrule
\end{tabular}}
\end{table}

\subparagraph{Results for non-informative observation processes}
When the observation process was not informative, all models estimated regression coefficients with null to negligible bias.
Coverage probability of the regression coefficients was also optimal, with slight under coverage for the intercept term \(\alpha_0\) and the treatment effect \(\alpha_1\) for estimates originating from the IIVW model.
Mean squared errors were similar across the range of scenarios with a non-informative observation process.
Bias for the variance of the residual error term was null to negligible as well, with good coverage.
Conversely, the variability of the random intercept \(v\) was estimated with slight negative bias from all models, with sub-par coverage (between 90\% and 95 \%); this is expected as we use maximum likelihood and not restricted maximum likelihood.
Finally, the estimated variance of the random effect linking the two outcomes in the joint model was positively biased with coverage of approximately 75\%; the magnitude of bias decreased as the baseline intensity \(\lambda\) increased.

\subparagraph{Results for informative observation processes}
When generating data from a \(\Gamma\) distribution depending on treatment only, all models were able to estimate the regression coefficients with no bias, optimal coverage probability, and comparable mean squared errors.
Conversely, in all other scenarios, the models performed quite differently.
In the scenario with observation times simulated from a \(\Gamma\) distribution depending on treatment and previous values of the longitudinal outcome, all models but model B (adjusting for the number of measurements) could estimate the treatment effect with null or minimal bias; model B overestimated the treatment effect.
The same pattern was observed for coverage of the treatment effect, with model B under-covering, and for the mean squared errors.
The effect of time was estimated with small bias and good coverage from all models, with model E (IIVW model) performing slightly worst; mean squared errors were comparable.
In scenarios simulated from a joint model, as expected, the joint model (model A) performed best overall, with minimal to no bias, optimal coverage, and the lowest mean squared errors.
Model C (adjusting for the cumulative number of measurements) and model D (plain mixed model) overestimated the intercept term and underestimated the treatment effect while showing small bias when estimating the effect of time.
Interestingly, both models showed that the bias when estimating the effect of time decreased as the baseline intensity \(\lambda\) increased: as expected, including more measurements allows to better estimate the effect of time.
Model B performed worst when estimating the effect of treatment, with large negative bias.
It also yielded biased intercept and effect of time, however, as with model C and D, bias for the estimate of time decreased as more measurements were available.
Finally, model E slightly overestimated the effect of treatment.
Model E showed increasing bias when estimating the intercept as the visiting process was denser, while (analogously as with model B, C, D) showed less biased estimates of the effect of time as the baseline intensity increased.
All models with the largest biases showed also poor coverage and the largest standard errors.
Overall, in settings simulated from a joint model, model B and model E performed worse and showed the largest biases.
In the scenario simulated from a joint model with a sparse observation process and regular yearly visits the joint model (model A) and the plain mixed model (model D) performed best, managing to recover the true values of all regression coefficients with no bias, and optimal coverage probabilities and mean squared errors.
Model B managed to estimate the effect of time with small bias, but largely overestimated the intercept and underestimated the treatment effect.
Model C managed to estimate the intercept and the treatment effect with small or no bias, but severely underestimated the effect of time.
Coverage and mean squared errors followed the same pattern.

\subparagraph{Results for the association parameter \(\gamma\)}

The estimating procedure worked well when the two sub models were not associated.
For instance, there was no bias, coverage probabilities were optimal, and mean squared errors were small - irrespectively of the baseline intensity of visit \(\lambda\).
Conversely, when the sub models were associated (\(\gamma = 1.50\)) the estimated association parameter was slightly negatively biased (\(-0.11\) to \(-0.06\)), with sub-optimal coverage (75\% to 83\%).
Mean squared error decreased when the baseline intensity of visit increased.
Finally, the scenario simulated from a joint model with a strong association parameter \(\gamma = 3.00\) and regular visits showed the worst performance, with large negative bias (-3.7289), poor coverage, and large mean squared error.
Including regular visits caused \(\gamma\) to shrink towards the null, with a median estimate of -0.7289.

\subparagraph{Convergence rates}
Convergence rates for all models included in this comparison were generally good.
All models showed a perfect convergence rate of 100\% except the joint model, which showed a lower convergence rate of 96\% and 99\% in two simulated scenarios, both with an informative observation process.
However, the remaining scenarios showed a perfect convergence rate for the joint model as well.

\section{Application}\label{section:application}

We fit the models included in this comparison to data obtained from the "Primary-Secondary Care Partnership to Prevent Adverse Outcomes in Chronic Kidney Disease" (PSP-CKD) study (ClinicalTrials.gov Identifier: NCT01688141, \citet{major_2019}).
PSP-CKD is a cluster randomised controlled pragmatic trial of enhanced chronic kidney disease (CKD) care against usual primary care management.
49 primary care practices from the Nene Clinical Commissioning Group, Northamptonshire, United Kingdom, were randomised to either enhanced care or usual care; informed consent was provided at the practice level.
Adult individuals with CKD were identified from each practice by using a research version of the web-based CKD management and audit tool IMPAKT (available at \url{http://www.impakt.org.uk/}); all data was anonymised prior to removal from the primary care practice.
Individuals were included if a recorded estimated glomerular filtration rate (eGFR) below 60 ml/min/1.73m\textsuperscript{2} was found during 5 years before the date of randomisation; eGFR was estimated using the MDRD equation \citep{levey_1999}.

We extracted baseline data (collected retrospectively at the date of randomisation and up to 5 years prior) from the PSP-CKD study consisting of all longitudinal eGFR measurements recorded during routine visits to the practices prior to randomisation; we also extracted the gender of each participant.
This resulted in 239,468 eGFR measurements for 36,527 individuals, of which 14,268 (39\%) were males and the remaining 22,259 (61\%) were females.
The median gap time between observations was 0.35 years (129 days), with inter-quartile interval (IQI) of 0.11 -- 0.74 years (39 -- 272 days).
We aim to evaluate whether the longitudinal eGFR trajectory before randomisation to treatment differs between males and females.

We start by evaluating whether the visiting process could be informative.
First, we computed the Spearman's rank correlation between gap time and gender: (\(\rho = 0.01\)).
The correlation coefficient was significantly different than zero.
Second, we fitted a linear mixed model for gap time versus gender with a random intercept and a random gender effect, and we found a significant association, as females had an 8.56-days longer gap time (95\% CI: 5.58 -- 11.54).
Finally, fitting the Andersen-Gill model for the observation process as described in Section \ref{section:iivw} with gender as the only covariate included in the model yielded a hazard ratio of 0.9589 (with 95\% C.I.: 0.9398 -- 0.9783) for females compared to males.
In conclusion, we found gap time to be associated with gender; hence, we deem the visiting process to likely be informative.

We fit the models included in the comparison, with gender as the binary exposure variable. The joint model included gender as the only covariate in the observation process submodel, and so did the recurrent events model utilised to fit weights for the IIVW model.

\begin{figure}
  \centering
  \includegraphics[width = 0.75\textwidth]{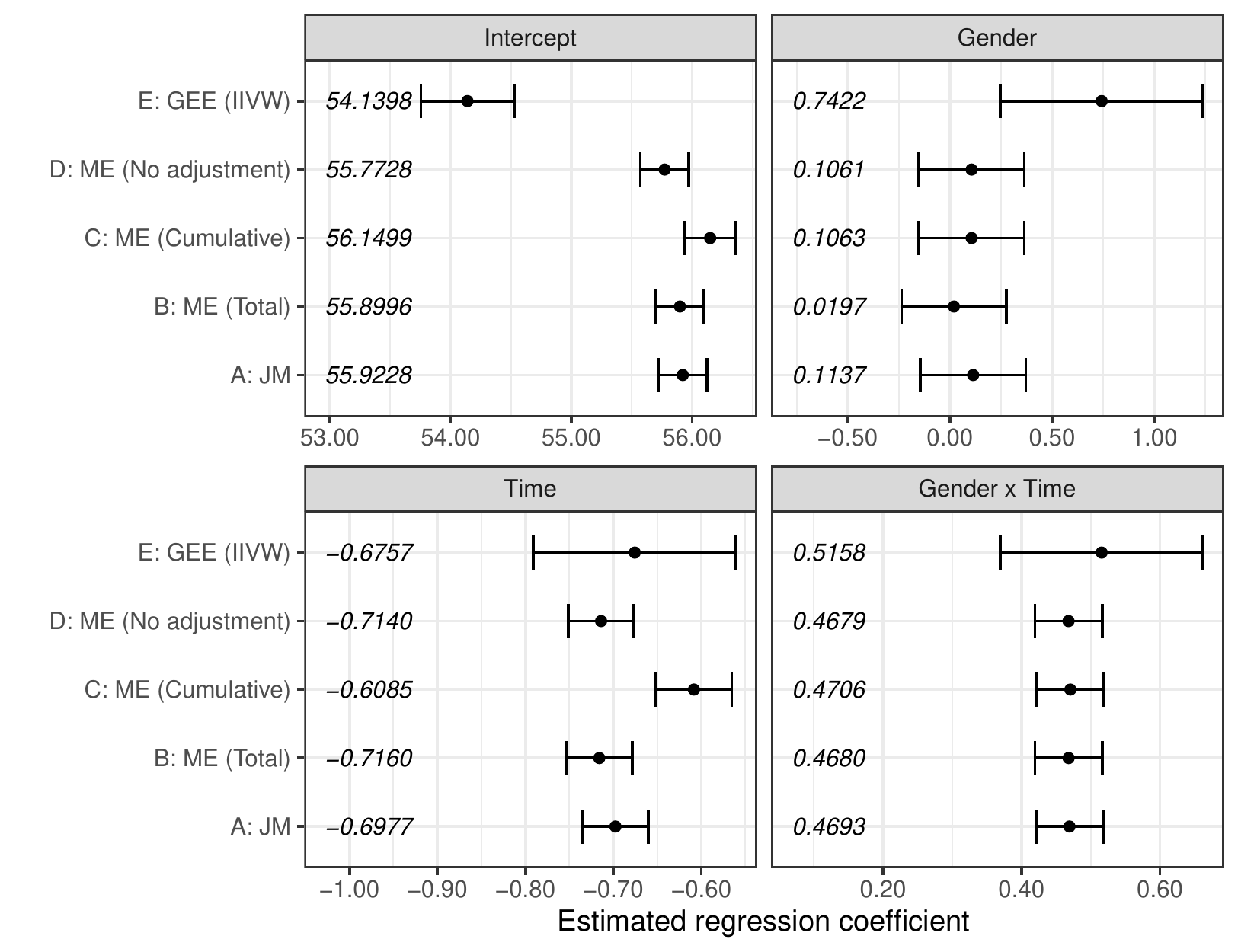}
  \caption{Forest plot with estimated coefficients for the longitudinal component, models fit to the application data from the PSP-CKD study. Each estimated coefficient is included as text placed on the leftmost side of each subplot.}
  \label{fig:ex-p2}
\end{figure}

The estimated coefficients for the longitudinal trajectory from each model are presented in Figure \ref{fig:ex-p2}.
The marginal model estimated an intercept and gender effect significantly different than the other four models: specifically, the estimated intercept from the marginal model was approximately 2 units lower, and the effect of gender was approximately 7 times higher and statistically significant, compared to a non-statistically significant effect of gender estimated by the remaining models.
The estimated effect of time was similar between all models (approximately -0.70 per unit of time), with the exception of the mixed model adjusting for the cumulative number of measurements as a time-varying covariate (estimated effect of approximately -0.60).
The interaction between gender and time was similarly estimated by all models, ranging between 0.4679 and 0.5158, and was statistically significant.
This showed that females had a slower decline in renal function over time compared to men.
The estimated coefficient for the observation process from the joint model shows a reduced risk of having a measured value for females compared to males (approximately 6\%, hazard ratio of 0.9417 with 95\% CI: 0.9245 -- 0.9589).
This value jointly with the estimated value of the association parameter \(\gamma\) (-3.8018, 95\% CI: -3.9943 to -3.6092) seem to confirm that the observation process is informed by gender.

\begin{figure}
  \centering
  \includegraphics[width = 0.75\textwidth]{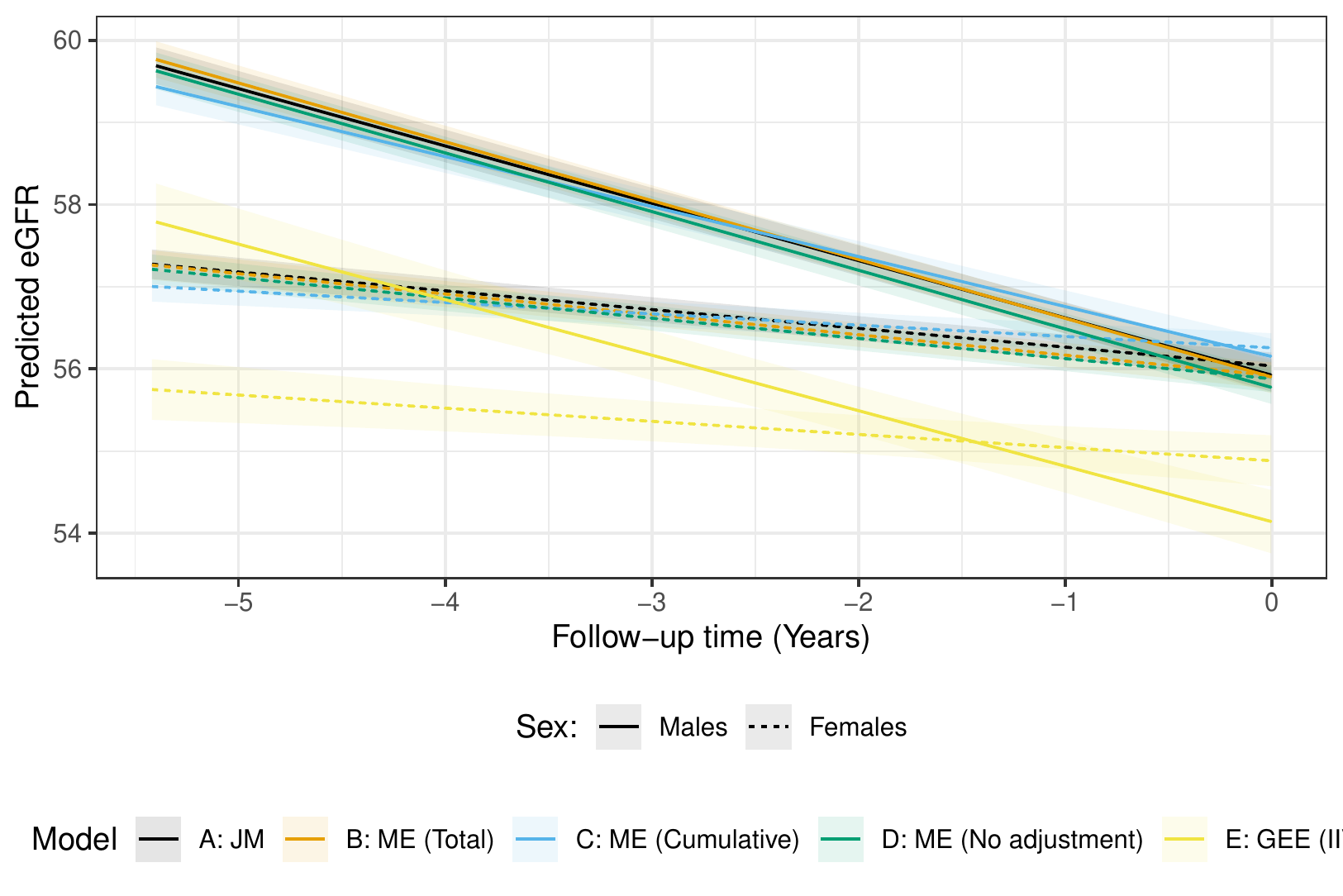}
  \caption{Predicted longitudinal trajectories from the models fit to the application data from the PSP-CKD study. The solid lines represent estimated trajectories for males, while the dashed lines represent trajectories for females. Colours identify the model.}
  \label{fig:ex-p1}
\end{figure}

Overall all models estimated a similar longitudinal trajectory (Figure \ref{fig:ex-p1}), with the IIVW model being the exception.
We saw in the results of our simulations in Section \ref{section:simulation} that the IIVW model yielded biased results for the exposure and the intercept of the longitudinal model under a variety of scenarios, and we observe this difference in our applied setting as well.
Interestingly, all other models performed similarly - even the mixed model adjusting for the total number of measurements; our simulations showed that the effect of a binary exposure was estimated with bias, but we did not saw this difference in practice.

\section{Discussion}\label{section:discussion}

In this article, we formalise the problem of informative visiting process within a framework of multivariate generalised linear and non-linear mixed effects models, including causal considerations.
Via Monte Carlo simulation, we illustrate (1) how ignoring an informative visiting process leads to biased estimates of the regression coefficient of a longitudinal model and (2) we compare some of the methods that have been proposed in the literature to account for it.
To the best of our knowledge, there is only one comparison currently in the literature \citep{neuhaus_2018}, albeit they include different models in their comparison and simulate an informative observation process differently by first generating a grid of potential observation times and then relating the probability of being observed to a given functional form of current (or lagged) covariates.
They also do not include a joint model analogous to the model introduced in our manuscript in Section \ref{section:model} in their comparison.

As expected, the joint model that accounts for the informative observation process by modelling it via a recurrent events survival model performed best.
Interestingly, the mixed effects model that disregarded completely the observation process performed worse than the joint model but outperformed other methods; the inflation in the variance of the random intercept of the plain mixed model seemed to capture part (if not most) of the variability due to the observation process, although this result needs to be thoroughly tested in more complex scenarios (e.g. with random effects of time, etc.).
The mixed models adjusting for the total number of measurements or the cumulative number of measurements (as a time-varying covariate) performed worst, and we would not recommend their usage in practice in these settings; this finding contrasts the findings of \citet{goldstein_2016}, although their settings were quite different than ours.
Further to that, they acknowledged the potential for collider bias (due to conditioning on a collider, the number of measurements) when the phenotyping algorithm for determining the exposure has high sensitivity; indeed, in our settings, the sensitivity is perfect as there is no misspecification of the exposure.
An additional possible explanation could be that in our settings the model adjusting for the total number of measurements is in fact conditioning on the future, as the total number of observations is not determined at the beginning of the study.
This may be explaining the poor performance of this method in the settings of our simulations.
The performance of the marginal model fitted using generalised estimating equations and inverse intensity of visit weights laid between the plain mixed model and the remaining mixed models; furthermore, its performance seemed to improve when the observation pattern became denser, except for the intercept term \(\alpha_0\).
This pattern was generally observed throughout all scenarios and models, as the performance seemed to increase with more frequent observation patterns; this finding is consistent with \citet{hernan_2009}.
The results of our simulations are consistent with those of \citet{neuhaus_2018}: the IIVW approach showed bias in all the settings of their simulation where the observation process was informative, even when adding regular visits to the study.
To compute the weights of the IIVW approach applied researchers need to correctly specify the model for the visit process, a challenging task - especially when not all the information required to fit the correctly specified model is observed (or known).
We also observed that the IIVW model performed quite differently than the other methods in our applied example, although the observed difference does not seem to be clinically relevant.

Most importantly, our simulations show that under the null all the approaches compared in this study produce unbiased estimates of the regression coefficients, the implication being that over modelling the observation process does not seem to introduce bias in the analysis.
In settings where it is not clear whether the observation process is informative or not, fitting the joint model would provide applied researchers with a method for estimating (and testing) the association between the two outcomes: this could be especially useful e.g. as a sensitivity analysis of standard mixed-effects models.

The joint model for the observation process and a longitudinal outcome that we described in Section \ref{section:model} can be further extended.
For instance, additional random effects could be introduced in the model to account for, say, heterogeneity in the trajectory of the longitudinal outcome over time.
The functional form of the effect of time (both fixed and random) could also be generalised by using fractional polynomials or splines; the longitudinal trajectories need to be modelled appropriately and best fit could be assessed via information criteria such as the AIC and BIC.
In fact, in the applied example of Section \ref{section:application} we assumed a linear effect of time on eGFR for simplicity; in actual applied projects one should assess whether the final model is correctly specified.
One could also extend the model to account for time-varying treatments, in both the observation process and longitudinal outcome sub models.
That would however require further investigations to assess the performance of the joint model in those settings.

We assumed the treatment to be constant over time for simplicity, but in real-life settings individuals are likely to start and drop treatment when deemed necessary by their treating physician.
We assumed the baseline hazard of the recurrent events model for the observation process to follow a Weibull distribution: this assumption could be further relaxed, and one could assume any parametric function, or even use flexible, spline-based formulations (e.g. \citet{royston_2002}).
Additionally, for diseases with a high mortality rate, a terminal event that truncates observation of the longitudinal process is likely to be informative in the sense that it likely correlates with disease severity.
That is, dropout is likely to be informative as the tendency to drop out after the occurrence of a terminal event is related to the current level of the longitudinally recorded biomarker.
The proposed model could be easily extended to include a third equation with a time to event submodel for the dropout process, as in \citet{liu_2008a}.
All of these extensions can be fit within the general framework of \citet{crowther_2017} using the Stata command \texttt{merlin}.
Finally, we could explore the association structure between the two sub models.
For instance, we could reverse the association structure and include \(\gamma\) in the observation submodel: in that setting, assuming a positive association, higher values of the longitudinal process would lead to a more frequent visiting process (and vice-versa in the setting of negative association).
The observation process could also depend on lagged values of the longitudinal outcome or of the exposure; this would relax the semi-Markov assumption in some of our data-generating mechanisms.
More biologically (and clinically) plausible association structures (such as the current value, current slope, cumulative effect parametrisations) could also be investigated; more details in \citet{rizopoulos_book}.

In conclusion, it is important to account for the visiting process when analysing health care utilisation data and we showed that ignoring it leads to biased estimates.
Given the wide range of applied settings in which this could be relevant, the review of \citet{farzanfar_2017} points towards a lack of awareness of the problem and the lack of readily available, user friendly software to fit more complex joint models; throughout this paper, we outlined a framework in which \texttt{merlin} could be easily used to fit complex joint model and help to reduce this translational gap.
We provide example code using Stata in the Online Supplementary Material.

\section*{Acknowledgements}

The authors would like to thank all primary care practices and Nene Clinical Commissioning Group for participating in the PSP-CKD study.
The PSP-CKD study was funded by the National Institute for Health Research (NIHR) Collaboration for Leadership in Applied Health Research and Care (CLAHRC) East Midlands.
Ongoing support for the study is funded by NIHR CLAHRC East Midlands and Kidney Research UK (Grant TF2/2015).
JKB is supported by the MRC Unit Programme number MC\_UU\_00002/5.
MJC is partially funded by the MRC-NIHR Methodology Research Panel (MR/P015433/1).

\bibliography{sim1}

\begin{thebibliography}{35}
\expandafter\ifx\csname natexlab\endcsname\relax\def\natexlab#1{#1}\fi
\expandafter\ifx\csname url\endcsname\relax
  \def\url#1{\texttt{#1}}\fi
\expandafter\ifx\csname urlprefix\endcsname\relax\def\urlprefix{URL: }\fi

\bibitem[{Andersen and Gill(1982)}]{andersen_1982}
Andersen, P.~K. and Gill, R.~D. (1982) Cox's regression model for counting
  processes: a large sample study.
\newblock \textit{The Annals of Statistics}, \textbf{10}, 1100--1120.

\bibitem[{Benchimol et~al.(2015)Benchimol, Smeeth, Guttmann, Harron, Moher,
  Petersen, S{\o}rensen, {von Elm}, Langan and {RECORD Working
  Committee}}]{benchimol_2015}
Benchimol, E.~I., Smeeth, L., Guttmann, A., Harron, K., Moher, D., Petersen,
  I., S{\o}rensen, H.~T., {von Elm}, E., Langan, S.~M. and {RECORD Working
  Committee} (2015) {The REporting of studies Conducted using Observational
  Routinely-collected health Data (RECORD) statement}.
\newblock \textit{PLoS medicine}, \textbf{12}, e1001885.

\bibitem[{Bender et~al.(2005)Bender, Augustin and Blettner}]{bender_2005}
Bender, R., Augustin, T. and Blettner, M. (2005) Generating survival times to
  simulate cox proportional hazards models.
\newblock \textit{Statistics in Medicine}, \textbf{24}, 1713--1723.

\bibitem[{B{\r{u}}{\v{z}}kov{\'a} et~al.(2010)B{\r{u}}{\v{z}}kov{\'a}, Brown
  and John-Stewart}]{buzkova_2010}
B{\r{u}}{\v{z}}kov{\'a}, P., Brown, E.~R. and John-Stewart, G.~C. (2010)
  Longitudinal data analysis for generalized linear models under
  participant-driven informative follow-up: an application in maternal health
  epidemiology.
\newblock \textit{American Journal of Epidemiology}, \textbf{171}, 189--197.

\bibitem[{B{\r{u}}{\v{z}}kov{\'a} and Lumley(2007)}]{buzkova_2007}
B{\r{u}}{\v{z}}kov{\'a}, P. and Lumley, T. (2007) Longitudinal data analysis
  for generalized linear models with follow-up dependent on outcome-related
  variables.
\newblock \textit{Canadian Journal of Statistics}, \textbf{35}, 485--500.

\bibitem[{Cole and Hern\'{a}n(2008)}]{cole_2008}
Cole, S.~R. and Hern\'{a}n, M.~A. (2008) Constructing inverse probability
  weights for marginal structural models.
\newblock \textit{American Journal of Epidemiology}, \textbf{168}, 656--664.

\bibitem[{Crowther(2017)}]{crowther_2017}
Crowther, M.~J. (2017) Extended multivariate generalised linear and non-linear
  mixed effects models.
\newblock \textit{arXiv preprint arXiv:1710.02223}.
\newblock \urlprefix\url{https://arxiv.org/abs/1710.02223}.

\bibitem[{Crowther(2018)}]{crowther_2018}
--- (2018) merlin - a unified modelling framework for data analysis and methods
  development in stata.
\newblock \textit{arXiv preprint arXiv:1806.01615}.
\newblock \urlprefix\url{https://arxiv.org/abs/1806.01615}.

\bibitem[{Crowther and Lambert(2012)}]{crowther_2012a}
Crowther, M.~J. and Lambert, P.~C. (2012) Simulating complex survival data.
\newblock \textit{The Stata Journal}, \textbf{12}, 674--687.

\bibitem[{Denaxas et~al.(2012)Denaxas, George, Herrett, Shah, Kalra, Hingorani,
  Kivimaki, Timmis, Smeeth and Hemingway}]{denaxas_2012}
Denaxas, S.~C., George, J., Herrett, E., Shah, A.~D., Kalra, D., Hingorani,
  A.~D., Kivimaki, M., Timmis, A.~D., Smeeth, L. and Hemingway, H. (2012) Data
  resource profile: cardiovascular disease research using linked bespoke
  studies and electronic health records ({CALIBER}).
\newblock \textit{International Journal of Epidemiology}, \textbf{41},
  1625--1638.

\bibitem[{Farzanfar et~al.(2017)Farzanfar, Abumuamar, Kim, Sirotich, Wang and
  Pullenayegum}]{farzanfar_2017}
Farzanfar, D., Abumuamar, A., Kim, J., Sirotich, E., Wang, Y. and Pullenayegum,
  E.~M. (2017) Longitudinal studies that use data collected as part of usual
  care risk reporting biased results: a systematic review.
\newblock \textit{BMC Medical Research Methodology}, \textbf{17}, 133.

\bibitem[{Gasparini(2018)}]{gasparini_2018}
Gasparini, A. (2018) {rsimsum: Summarise results from Monte Carlo simulation
  studies}.
\newblock \textit{Journal of Open Source Software}, \textbf{3}, 739.

\bibitem[{Goldstein et~al.(2016)Goldstein, Bhavsar, Phelan and
  Pencina}]{goldstein_2016}
Goldstein, B.~A., Bhavsar, N.~A., Phelan, M. and Pencina, M.~J. (2016)
  Controlling for informed presence bias due to the number of health encounters
  in an electronic health record.
\newblock \textit{American Journal of Epidemiology}, \textbf{184}, 847--855.

\bibitem[{Gruger et~al.(1991)Gruger, Kay and Schumacher}]{gruger_1991}
Gruger, J., Kay, R. and Schumacher, M. (1991) The validity of inferences based
  on incomplete observations in disease state models.
\newblock \textit{Biometrics}, \textbf{47}, 595--605.

\bibitem[{Hemmelgarn et~al.(2009)Hemmelgarn, Clement, Manns, Klarenbach, James,
  Ravani, Pannu, Ahmed, {MacRae} and {Scott-Douglas}}]{hemmelgarn_2009}
Hemmelgarn, B.~R., Clement, F., Manns, B.~J., Klarenbach, S., James, M.~T.,
  Ravani, P., Pannu, N., Ahmed, S.~B., {MacRae}, J. and {Scott-Douglas}, N.
  (2009) {Overview of the Alberta Kidney Disease Network}.
\newblock \textit{BMC Nephrology}, \textbf{10}, 30.

\bibitem[{Hern{\'a}n et~al.(2004)Hern{\'a}n, {Hern{\'a}ndez-D{\'i}az} and
  Robin}]{hernan_2004}
Hern{\'a}n, M.~A., {Hern{\'a}ndez-D{\'i}az}, S. and Robin, J.~M. (2004) A
  structural approach to selection bias.
\newblock \textit{Epidemiology}, \textbf{15}, 615--625.

\bibitem[{Hern{\'a}n et~al.(2009)Hern{\'a}n, McAdams, McGrath, Lanoy and
  Costagliola}]{hernan_2009}
Hern{\'a}n, M.~A., McAdams, M., McGrath, N., Lanoy, E. and Costagliola, D.
  (2009) Observation plans in longitudinal studies with time-varying
  treatments.
\newblock \textit{Statistical Methods in Medical Research}, \textbf{18},
  27--52.

\bibitem[{Kurland et~al.(2009)Kurland, Johnson, Egleston and
  Diehr}]{kurland_2009}
Kurland, B.~F., Johnson, L.~L., Egleston, B.~L. and Diehr, P.~H. (2009)
  Longitudinal data with follow-up truncated by death: match the analysis
  method to research aims.
\newblock \textit{Statistical Science}, \textbf{24}, 211.

\bibitem[{Levey et~al.(1999)Levey, Bosch, {Breyer Lewis}, Greene, Rogers, Roth
  and {The Modification of Diet in Renal Disease Study Group}}]{levey_1999}
Levey, A.~S., Bosch, J.~P., {Breyer Lewis}, J., Greene, T., Rogers, N., Roth,
  D. and {The Modification of Diet in Renal Disease Study Group} (1999) A more
  accurate method to estimate glomerular filtration rate from serum creatinine:
  A new prediction equation.
\newblock \textit{Annals of Internal Medicine}, \textbf{130}, 461--470.

\bibitem[{Liu et~al.(2008)Liu, Huang and {O'Quigley}}]{liu_2008a}
Liu, L., Huang, X. and {O'Quigley}, J. (2008) Analysis of longitudinal data in
  presence of informative observational times and a dependent terminal event,
  with application to medical cost data.
\newblock \textit{Biometrics}, \textbf{64}, 950--958.

\bibitem[{Liu et~al.(2018)Liu, Zheng and Kang}]{liu_2018}
Liu, L., Zheng, C. and Kang, J. (2018) Exploring causality mechanism in the
  joint analysis of longitudinal and survival data.
\newblock \textit{Statistics in Medicine}, \textbf{37}, 3733--3744.

\bibitem[{Major et~al.(2019)Major, Brown, Shepherd, Rogers, Pickering, Warwick,
  Barber, Ashra, Morris and Brunskill}]{major_2019}
Major, R.~W., Brown, C., Shepherd, D., Rogers, S., Pickering, W., Warwick,
  G.~L., Barber, S., Ashra, N.~B., Morris, T. and Brunskill, N.~J. (2019) The
  primary-secondary care partnership to improve outcomes in chronic kidney
  disease ({PSP-CKD}) study: A cluster randomized trial in primary care.
\newblock \textit{Journal of the American Society of Nephrology}.

\bibitem[{McCulloch et~al.(2016)McCulloch, Neuhaus and Olin}]{mcculloch_2016}
McCulloch, C.~E., Neuhaus, J.~M. and Olin, R.~L. (2016) Biased and unbiased
  estimation in longitudinal studies with informative visit processes.
\newblock \textit{Biometrics}, \textbf{72}, 1315--1324.

\bibitem[{Morris et~al.(2019)Morris, White and Crowther}]{morris_2019}
Morris, T.~P., White, I.~R. and Crowther, M.~J. (2019) Using simulation studies
  to evaluate statistical methods.
\newblock \textit{Statistics in Medicine}, \textbf{38}, 2074--2102.

\bibitem[{Neuhaus et~al.(2018)Neuhaus, {McCulloch} and Boylan}]{neuhaus_2018}
Neuhaus, J.~M., {McCulloch}, C.~E. and Boylan, R.~D. (2018) Analysis of
  longitudinal data from outcome-dependent visit processes: Failure of proposed
  methods in realistic settings and potential improvements.
\newblock \textit{Statistics in Medicine}, \textbf{37}, 4457--4471.
\newblock \urlprefix\url{https://doi.org/10.1002%2Fsim.7932}.

\bibitem[{Pinheiro and Bates(1995)}]{pinheiro_1995}
Pinheiro, J.~C. and Bates, D.~M. (1995) Approximations to the log-likelihood
  function in the nonlinear mixed-effects model.
\newblock \textit{Journal of Computational and Graphical Statistics},
  \textbf{4}, 12--35.

\bibitem[{Pullenayegum and Lim(2016)}]{pullenayegum_2016}
Pullenayegum, E.~M. and Lim, L.~S. (2016) Longitudinal data subject to
  irregular observation: A review of methods with a focus on visit processes,
  assumptions, and study design.
\newblock \textit{Statistical Methods in Medical Research}, \textbf{25},
  2992--3014.

\bibitem[{{R Core Team}(2019)}]{Rmanual}
{R Core Team} (2019) \textit{R: A Language and Environment for Statistical
  Computing}.
\newblock R Foundation for Statistical Computing, Vienna, Austria.
\newblock \urlprefix\url{https://www.R-project.org/}.

\bibitem[{Rizopoulos(2012)}]{rizopoulos_book}
Rizopoulos, D. (2012) \textit{Joint Models for Longitudinal and Time-to-Event
  Data: With Applications in {R}}.
\newblock Chapman and Hall / CRC Biostatistics Series. Chapman and Hall / CRC.

\bibitem[{Robins et~al.(1995)Robins, Rotnitzky and Zhao}]{robins_1995}
Robins, J.~M., Rotnitzky, A. and Zhao, L.~P. (1995) Analysis of semiparametric
  regression models for repeated outcomes in the presence of missing data.
\newblock \textit{Journal of the American Statistical Association},
  \textbf{90}, 106--121.

\bibitem[{Royston and Parmar(2002)}]{royston_2002}
Royston, P. and Parmar, M.~K. (2002) Flexible parametric proportional-hazards
  and proportional-odds models for censored survival data, with application to
  prognostic modelling and estimation of treatment effects.
\newblock \textit{Statistics in Medicine}, \textbf{21}, 2175--2197.

\bibitem[{Runesson et~al.(2015)Runesson, Gasparini, Qureshi, Norin, Evans,
  Barany, Wettermark, Elinder and Carrero}]{runesson_2015}
Runesson, B., Gasparini, A., Qureshi, A.~R., Norin, O., Evans, M., Barany, P.,
  Wettermark, B., Elinder, C.~G. and Carrero, J.~J. (2015) {The Stockholm
  CREAtinine Measurements (SCREAM) project: protocol overview and regional
  representativeness}.
\newblock \textit{Clinical Kidney Journal}, \textbf{9}, 119--127.

\bibitem[{Tanuseputro et~al.(2015)Tanuseputro, Wodchis, Fowler, Walker, Bai,
  Bronskill and Manuel}]{tanuseputro_2015}
Tanuseputro, P., Wodchis, W.~P., Fowler, R., Walker, P., Bai, Y.~Q., Bronskill,
  S.~E. and Manuel, D. (2015) The health care cost of dying: a population-based
  retrospective cohort study of the last year of life in {Ontario, Canada}.
\newblock \textit{PLoS One}, \textbf{10}, e0121759.

\bibitem[{{Van Ness} et~al.(2009){Van Ness}, Allore, Fried and
  Lin}]{van_ness_2009}
{Van Ness}, P.~H., Allore, H.~G., Fried, T.~R. and Lin, H. (2009) Inverse
  intensity weighting in generalized linear models as an option for analyzing
  longitudinal data with triggered observations.
\newblock \textit{American Journal of Epidemiology}, \textbf{171}, 105--112.

\bibitem[{Wu and Carroll(1988)}]{wu_1988}
Wu, M.~C. and Carroll, R.~J. (1988) Estimation and comparison of changes in the
  presence of informative right censoring by modeling the censoring process.
\newblock \textit{Biometrics}, \textbf{44}, 175--188.

\end{thebibliography}

\end{document}